\documentclass[aps,prl,showpacs,twocolumn,groupedaddress]{revtex4}

\usepackage{graphicx}
\usepackage{dcolumn}
\usepackage{bm}

\begin{document}

\preprint{APS/123-QED}

\title{Distorted Kagome Lattice Generated by a Unique Orbital Arrangement in the Copper Mineral KCu$_3$As$_2$O$_7$(OH)$_3$}

%

\author{Yoshihiko Okamoto, Hajime Ishikawa, G\o ran J. Nilsen, and Zenji Hiroi}
\affiliation{
Institute for Solid State Physics, University of Tokyo, Kashiwa 277-8581, Japan
}

\date{\today}

\begin{abstract}
We study polycrystalline samples of KCu$_3$As$_2$O$_7$(OH)$_3$, a new candidate spin-1/2 kagome antiferromagnet, by magnetic susceptibility and heat capacity measurements above 2 K. The unique arrangement of the 3$z^2 - r^2$ and $x^2 - y^2$ orbitals on the Cu$^{2+}$ kagome net is noted and compared to the orbital patterns found in other kagome minerals. It is suggested that this orbital arrangement gives rise to one antiferromagnetic and two ferromagnetic interactions on isosceles triangles forming a highly distorted kagome lattice. KCu$_3$As$_2$O$_7$(OH)$_3$ is found to show an antiferromagnetic long-range order at $T_{\mathrm{N}}$ = 7.2 K. Remarkably, a spin entropy is more gradually released upon cooling below $T_{\mathrm{N}}$ compared with a conventional magnetic long-range order, which may originate from the geometrical frustration still present in this highly distorted kagome lattice. 
\end{abstract}

\pacs{Valid PACS appear here}
\maketitle

The spin-1/2 kagome antiferromagnet (KAFM) is one of the most intriguing systems in the field of magnetism, owing to the strong frustration that arises when regular triangles are connected in a corner-sharing geometry. It is theoretically expected that the spin-1/2 KAFM is a candidate for showing a spin liquid ground state instead of a classical ordered state owing to its strong frustration and strong quantum fluctuation~\cite{BalentsReview}. However, the nature of the ground state is still under debate. Experimentally, two Cu minerals, herbertsmithite~\cite{Herb1} and vesignieite~\cite{Vesi1}, have been extensively studied as model compounds. However, their intrinsic ground states are still unclear, because they suffer from certain perturbations caused by lattice defects, disorder, or low crystallinity\cite{Herb3, Vesi1}. 

Recently, the effects of distortion in the magnetic interactions on the kagome lattice have come a topic of great interest, largely because intriguing magnetic properties were discovered in the distorted kagome Cu mineral volborthite~\cite{Vol1,Vol2,Vol3,Vol4}. Expanding or contracting the triangles making up the ideal kagome lattice along one side gives rise to a distorted kagome lattice comprising isosceles triangles with two different magnetic interactions $J$ and $J^{\prime}$, as shown in Fig. 1(a). Theoretical studies suggest that the ground state of such distorted KAFMs is not far from that of undistorted one~\cite{DistKagome2}, because frustration still remains although the degeneracy is partly lifted. Moreover, in a small magnetic field, a magnetically ordered state is expected to appear even for a tiny distortion of a few percent~\cite{Kaneko}. On the other hand, for a highly distorted KAFM with $J$ much larger than $J^{\prime}$, the one-dimensional bosonization method and renormalization group analysis showed that a peculiar helical ordered state appears at zero magnetic field instead of a spin liquid, where the magnitude of magnetic moments of Cu$^{2+}$ spins differs significantly between two Cu sites~\cite{DistKagome3, DistKagome1}. 

In Cu minerals with kagome lattices, the pattern of the Cu$^{2+}$ orbital arrangement determines whether the kagome lattice is distorted. Since the Cu$^{2+}$ ion is always in a heavily deformed octahedral coordination owing to the Jahn-Teller effect, either the $x^2 - y^2$ or 3$z^2 - r^2$ orbital is selected, depending on the sign of the deformation~\cite{HiroiReview}. When the Cu-anion distances for the four equatorial anions are considerably shorter than those for the two axial anions, the $x^2 - y^2$ orbital carries an $S$ = 1/2 spin. In contrast, when two axial anions come closer to the Cu ion with four equatorial anions further apart, the 3$z^2 - r^2$ orbital carries an $S$ = 1/2 spin. Hence, we can easily identify which orbital carries a spin from the shape of the octahedra in copper minerals. 

Herbertsmithite, vesignieite, and volborthite, all mentioned above, exhibit different patterns of orbital arrangement, as shown in Figs. 1(b)-1(d), respectively. In herbertsmithite, the $x^2 - y^2$ orbital carries a spin on every Cu$^{2+}$ ion: the Cu$^{2+}$ ion is coordinated by an (OH)$_4$Cl$_2$ octahedron with the two Cl$^-$ ions at the axial positions~\cite{Herb1}. Since the Cu-Cl distance (2.770 \r{A}) is much longer than the Cu-O distance (1.982 \r{A}), the $x^2 - y^2$ orbital is selected and arranged so as to keep a three-fold rotation axis at the center of a Cu$_3$ triangle, resulting in an undistorted kagome lattice with $J$. In vesignieite, the Cu$^{2+}$ ion is coordinated by an O$_4$(OH)$_2$ octahedron, where the two OH$^-$ ions occupy the axial positions, as shown in Fig. 1(c)~\cite{Vesi0, Vesi2}. In contrast to that in herbertsmithite, the 3$z^2 - r^2$ orbital carries a spin, because the axial Cu-OH distance (1.91 \r{A}) is much shorter than the equatorial Cu-O distances (2.17-2.18 \r{A}). This gives rise to an almost undistorted kagome lattice similar to that seen in herbertsmithite. On the other hand, the orbital pattern of volborthite is substantially different: the 3$z^2 - r^2$ and $x^2 - y^2$ orbitals are selected in the Cu1 and Cu2 sites, respectively~\cite{HiroiReview}, as shown in Fig. 1(d), resulting in a distorted kagome lattice with apparently different magnetic interactions between neighboring Cu2 spins ($J$) and Cu1-Cu2 spins ($J^{\prime}$). However, the signs and magnitudes of $J$ and $J^{\prime}$ are still under debate~\cite{VolTheo2}. 

\begin{figure}
\begin{center}
\includegraphics[width=7.2cm]{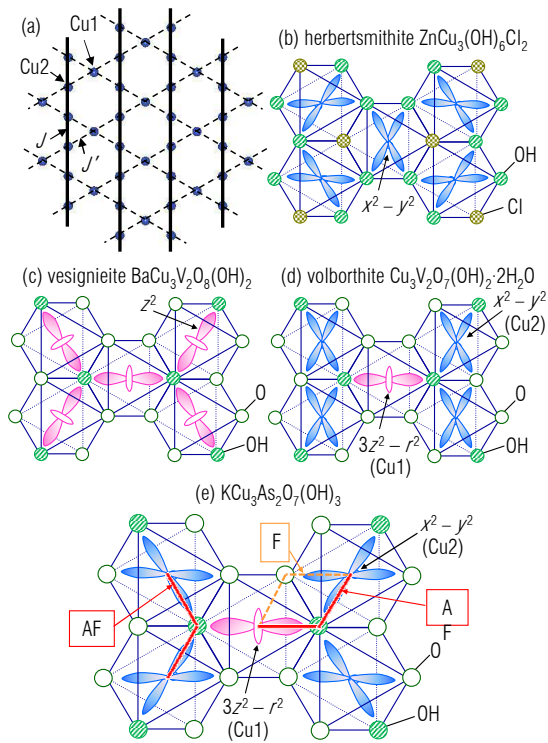}
\caption{(Color online) (a) Distorted kagome lattice made of two kinds of Cu ions, each carrying an $S$ = 1/2 spin. Magnetic interactions, $J$ and $J^{\prime}$, for Cu2-Cu2 and Cu1-Cu2 are represented by a thick solid line and a thin broken line, respectively. Arrangements of the Cu 3$d$ orbitals within the kagome layer are depicted for herbertsmithite (b), vesignieite (c), volborthite (d), and KCu$_3$As$_2$O$_7$(OH)$_3$ (e). For KCu$_3$As$_2$O$_7$(OH)$_3$, all possible nearest-neighbor superexchange pathways are shown: the thick solid lines represent Cu-OH-Cu pathways giving antiferromagnetic interactions, and the thick broken line represents a Cu-O-Cu pathway giving a ferromagnetic interaction.}
\label{F1}
\end{center}
\end{figure}

As described above, there is a clear relation between patterns of orbital arrangement and magnetic interactions in kagome Cu minerals. To clarify this relation is essential for understanding the magnetism of kagome lattices. In this letter, we report the copper mineral KCu$_3$As$_2$O$_7$(OH)$_3$, which has not been focused on thus far, but shows a novel spin-1/2 distorted kagome lattice. This compound was first prepared by Effenberger in 1989, but its physical properties have not been reported until now~\cite{KCuAs}. It crystallizes in a monoclinic structure with the space group of $C$2/$m$, as shown in Fig. 2, which is isomorphic to that of vesignieite. Compared with that of vesignieite, however, the orbital arrangement of KCu$_3$As$_2$O$_7$(OH)$_3$ is quite different: the 3$z^2 - r^2$ orbital is selected in the Cu1 site with axial Cu-OH distances $\sim$15\% shorter than the other four distances, while the $x^2 - y^2$ orbital is selected in the Cu2 site with axial Cu-O distances $\sim$20\% longer, as shown in Fig. 1(e)~\cite{KCuAs}. The difference in the orbital arrangement between KCu$_3$As$_2$O$_7$(OH)$_3$ and vesignieite may originate from the difference in size between the tetrahedra located above and below the hexagons in the kagome lattice: AsO$_4$ in KCu$_3$As$_2$O$_7$(OH)$_3$ is larger than VO$_4$ in vesignieite. The orbital arrangement in KCu$_3$As$_2$O$_7$(OH)$_3$ shares the orbital selection in the Cu1 and Cu2 sites with volborthite shown in Fig. 1(d), but the $x^2 - y^2$ orbitals are aligned in a staggered way in KCu$_3$As$_2$O$_7$(OH)$_3$. It is interesting to study how magnetic interactions are affected by this difference, and what kind of magnetic ground state results in the As compound. We have succeeded in preparing a single-phase sample of KCu$_3$As$_2$O$_7$(OH)$_3$ and studied its physical properties. KCu$_3$As$_2$O$_7$(OH)$_3$ is found to exhibit an antiferromagnetic long-range order at $T_{\mathrm{N}}$ = 7.2 K, which is peculiar because spin entropy is more gradually released upon cooling below $T_{\mathrm{N}}$ compared with that in the case of conventional magnetic transition.
The geometrical frustration that remains on the highly distorted kagome lattice may be important.

\begin{figure}
\begin{center}
\includegraphics[width=7.3cm]{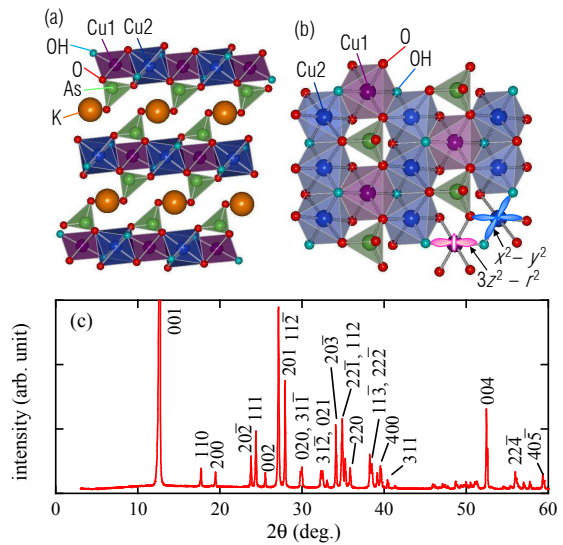}
\caption{(Color online) Crystal structure of KCu$_3$As$_2$O$_7$(OH)$_3$ viewed along the $b$ axis (a) and perpendicular to the $ab$ plane (b). The 3$z^2 - r^2$ and $x^2 - y^2$ orbitals selected at the Cu1 and Cu2 sites, respectively, are depicted on the lower right of (b). (c) XRD pattern from polycrystalline sample of KCu$_3$As$_2$O$_7$(OH)$_3$ taken at room temperature. Peak indices are given using a monoclinic unit cell with the lattice parameters $a$ = 12.299(2) \r{A}, $b$ = 5.9881(9) \r{A}, $c$ = 7.885(2) \r{A}, and $\beta$ = 117.86(14)$^{\circ}$.}
\label{F2}
\end{center}
\end{figure}

We have prepared a KCu$_3$As$_2$O$_7$(OH)$_3$ powder sample by the hydrothermal method. Starting materials, i.e., 2 g of KH$_2$AsO$_4$, 0.2 g of Cu(OH)$_2$, and 15 ml of 0.2 M KOH aqueous solution, were put in a Teflon beaker placed in a stainless-steel vessel. The vessel was sealed, heated at 220$^{\circ}$C for 24 h, and furnace-cooled to room temperature. The pale-green powder thus obtained was rinsed with water several times and dried at room temperature. Sample characterization was performed by powder X-ray diffraction (XRD) analysis with Cu-K$\alpha$ radiation at room temperature, employing a RINT-2000 diffractometer (Rigaku). All diffraction peaks observed in the powder XRD pattern shown in Fig. 2(c) are sharp, indicating good crystallinity. Moreover, they can be indexed on the basis of a monoclinic structure with lattice constants $a$ = 12.299(2) \r{A}, $b$ = 5.9881(9) \r{A}, $c$ = 7.885(2) \r{A}, and $\beta$ = 117.86(14)$^{\circ}$, close to those reported by Effenberger~\cite{KCuAs}. Magnetic susceptibility and heat capacity measurements above 2 K were performed in a MPMS and a PPMS (both Quantum Design), respectively.

\begin{figure}
\begin{center}
\includegraphics[width=7.4cm]{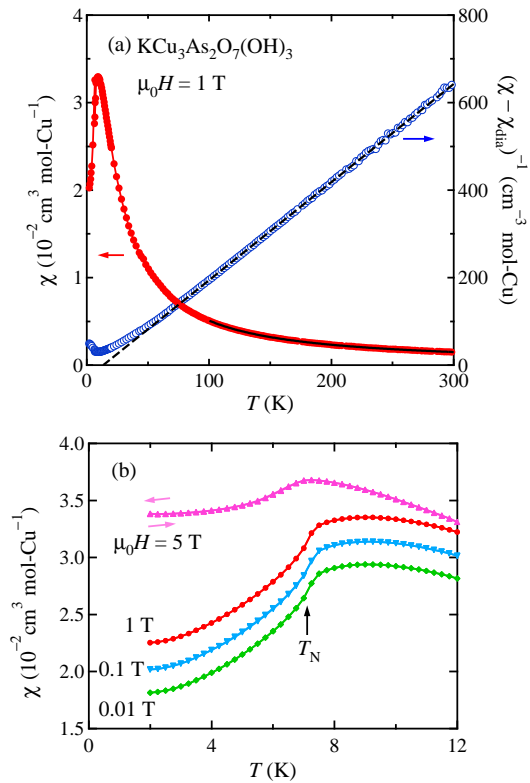}
\caption{(Color online) (a) Temperature dependence of magnetic susceptibility $\chi$ of a powder sample of KCu$_3$As$_2$O$_7$(OH)$_3$ at a magnetic field of 1 T. Inverse susceptibility is also shown in (a) after the subtraction of diamagnetic contributions from core electrons, $\chi_{\mathrm{dia}}$ = $-$5.93 $\times$ 10$^{-5}$ cm$^3$ mol-Cu$^{-1}$.~\cite{xdia} The solid curve and broken line show a fit to the Curie-Weiss law, $\chi$ = $C$/($T - \theta_{\mathrm{W}}$) $+$ $\chi_{\mathrm{dia}}$, which yields $C$ = 0.4458(5) cm$^3$ K mol-Cu$^{-1}$ and $\theta_{\mathrm{W}}$ = 13.4(12) K. (b) Temperature dependence of field-cooled and zero-field-cooled $\chi$ measured at magnetic fields of 0.01, 0.1, 1, and 5 T. For clarity, the 0.1 and 0.01 T data are shown with downward shifts by offsets of 2 $\times$ 10$^{-3}$ and 4 $\times$ 10$^{-3}$ cm$^3$ mol-Cu$^{-1}$, respectively. An antiferromagnetic ordering transition takes place at $T_{\mathrm{N}}$ = 7.2 K at $\mu_0H$ = 0.01 T, as shown by the arrow.}
\label{F3}
\end{center}
\end{figure}

Figure 3(a) shows the temperature dependence of the magnetic susceptibility $\chi$ measured on a polycrystalline sample of KCu$_3$As$_2$O$_7$(OH)$_3$. The inverse of ($\chi - \chi_{\mathrm{dia}}$), where $\chi_{\mathrm{dia}}$ is the diamagnetic contribution of core electrons ($\chi_{\mathrm{dia}}$ = $-$5.93 $\times$ 10$^{-5}$ cm$^3$ mol-Cu$^{-1}$)~\cite{xdia}, exhibits a linear dependence at high temperatures above $\sim$100 K. This indicates that the Curie-Weiss law is obeyed and that the Van Vleck contribution is negligible, as is always the case for copper compounds.~\cite{Herb1,Vesi1,Vol1}
A fit to the equation $\chi - \chi_{\mathrm{dia}}$ = $C$/($T - \theta_{\mathrm{W}}$), where $C$ and $\theta_{\mathrm{W}}$ are the Curie constant and Weiss temperature, respectively, yields $C$ = 0.4458(5) cm$^3$ K mol-Cu$^{-1}$ and $\theta_{\mathrm{W}}$ = 13.4(12) K. The $C$ corresponds to an effective moment of $\mu_{\mathrm{eff}}$ = 1.89 $\mu_{\mathrm{B}}$ per Cu atom, which gives a Lande g-factor of $g$ = 2.18 for $S$ = 1/2, slightly larger than 2; the $g$ is close to $g$ = 2.15 for volborthite and $g$ = 2.14 for vesignieite, as determined by ESR measurements~\cite{ESRvol, ESRvesi}. On the other hand, the small positive $\theta_{\mathrm{W}}$ indicates that the average magnetic interaction is weakly ferromagnetic, which is in contrast to the large negative $\theta_{\mathrm{W}}$'s of $-$300, $-$77, and $-$115 K for herbertsmithite, vesignieite, and volborthite, respectively~\cite{Herb1, Vol1, Vesi1}. Note, however, that the inverse susceptibility in Fig. 3(a) deviates upward from the Curie-Weiss line below $\sim$50 K. This implies that an antiferromagnetic interaction of this energy scale exists in KCu$_3$As$_2$O$_7$(OH)$_3$. Presumably, as later described in detail, ferromagnetic and antiferromagnetic interactions coexist and partially cancel each other, resulting in a small positive $\theta_{\mathrm{W}}$. 

\begin{figure}
\begin{center}
\includegraphics[width=7cm]{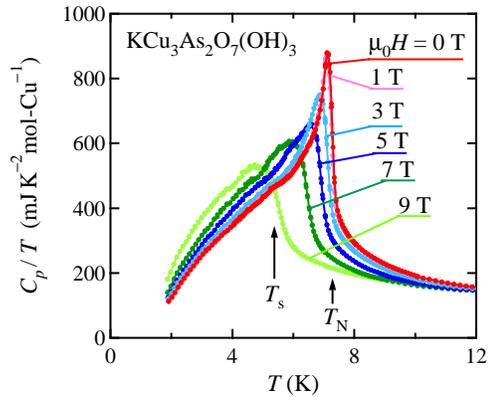}
\caption{(Color online) Temperature dependence of heat capacity divided by temperature of powder sample of KCu$_3$As$_2$O$_7$(OH)$_3$ measured at various fields from 0 to 9 T. $T_{\mathrm{N}}$ and $T_{\mathrm{s}}$ at zero magnetic field are also marked by arrows.}
\label{F4}
\end{center}
\end{figure}

Magnetic susceptibilities at temperatures below 12 K and measured at various magnetic fields up to 5 T are shown in Fig. 3(b). A broad peak at $\sim$9 K and a distinct kink at 7.2 K are observed in the low-field data. The former represents the onset of short-range magnetic correlations in a two-dimensional spin system, and the latter indicates a three-dimensional long range antiferromagnetic order. The order is not accompanied by a weak ferromagnetic moment, as sometimes observed in other kagome compounds. Moreover, there is no glassy component: the field and zero-field cooling curves completely overlap even at a small magnetic field of 0.01 T, as shown in Fig. 3(b). Note that the low-field data at 0.01, 0.1, and 1 T completely overlap each other (the latter two data are shown with offsets in Fig. 3(b)), while the 5 T data is substantially different from them. There may be a subtle change in magnetic structure at high magnetic fields.

The presence of a phase transition at $T_{\mathrm{N}}$ is also clearly evidenced in the heat capacity divided by temperature shown in Fig. 4. A sharp peak appears at 7.2 K at zero magnetic field, which agrees well with the kink temperature in the magnetic susceptibility curve measured at a magnetic field of 0.01 T. This peak in heat capacity moves to lower temperatures with increasing magnetic field, just as observed for the kink temperature in magnetic susceptibility, 
which is consistent with the antiferromagnetic nature of this order. There is no anomaly in heat capacity corresponding to the broad peak at 9 K in $\chi$, although $C_p$/$T$ begins to grow further above $T_{\mathrm{N}}$. These two features in $\chi$ and $C_p$ mean that short-range antiferromagnetic correlations are already developed above $T_{\mathrm{N}}$. 

Interestingly, the temperature dependence of $C_p$/$T$ below $T_{\mathrm{N}}$ shows an unusual feature. It does not exhibit a power law behavior as expected for a conventional N\'{e}el order, but rather a shoulder at $T_{\mathrm{s}}$ = 5.5 K at zero magnetic field. $T_{\mathrm{s}}$ shifts to lower temperatures with increasing magnetic field. Since there is no anomaly corresponding to $T_{\mathrm{s}}$ in the $\chi$ shown in Fig. 3(b), another magnetic phase transition is unlikely to occur at $T_{\mathrm{s}}$. This broad shoulder suggests that spins are not fully ordered at $T_{\mathrm{N}}$ and gradually release their remaining entropy at around $T_{\mathrm{s}}$. Similar entropy-releasing phenomena have been observed in classical triangular-lattice antiferromagnets such as 
NaCrO$_2$~\cite{NaCrO2}, where they were ascribed to geometrical frustration in the triangular lattices. The broad shoulder in $C_p$/$T$ in KCu$_3$As$_2$O$_7$(OH)$_3$ can also be due to the geometrical frustration of the distorted kagome lattice, though details still remain unclear. This interesting possibility will be clarified by future NMR or neutron scattering experiments. 

The magnetic interactions between Cu$^{2+}$ spins in KCu$_3$As$_2$O$_7$(OH)$_3$ must differ from those in vesignieite in spite of the similarity in crystal structure (Fig. 1); common to both, the Cu$^{2+}$ ions are coordinated by an O$_4$(OH)$_2$ octahedron, with the OH$^-$ and O$^{2-}$ ions occupying the axial and equilateral positions, respectively. In vesignieite, the superexchange interactions occur via the OH$^-$ ion, because all the 3$z^2 - r^2$ orbitals carrying spins point to the OH$^-$ ion above or below the Cu$_3$ triangle, as shown in Fig. 2(c); there are no apparent superexchange pathways between second-nearest neighbors in vesignieite. The Cu1-OH-Cu2 and Cu2-OH-Cu2 angles are 101.76$^{\circ}$ and 101.72$^{\circ}$, respectively~\cite{Vesi2}, resulting in significant antiferromagnetic interactions of almost identical magnitudes. Fits to magnetic susceptibility using a high-temperature series expansion for the isotropic kagome lattice yields $J$/$k_{\mathrm{B}}$ = 55 K in average~\cite{Vesi1}. 

For the magnetic interaction between Cu2 spins in KCu$_3$As$_2$O$_7$(OH)$_3$, a superexchange interaction via a Cu2-OH-Cu2 pathway must be crucial, because the lobes of two adjacent $x^2 - y^2$ orbitals point to the OH$^-$ ion above or below the center of the Cu$_3$ triangle, as shown in Fig. 1 (e). This superexchange interaction should be antiferromagnetic as in vesignieite: the Cu2-OH-Cu2 angle is 101.30$^{\circ}$. On the other hand, the magnetic interactions between Cu1 and Cu2 spins are more complicated, because different orbitals are selected. It is intuitive to divide the 3$z^2 - r^2$ orbital in the Cu1 site into two $x^2 - y^2$ type orbitals, i.e. the $z^2 - x^2$ and $z^2 - y^2$ orbitals: 3$z^2 - r^2$ = ($z^2 - x^2$) + ($z^2 - y^2$). Then, superexchange interactions between a Cu1 spin and each of two neighboring Cu2 spins occur between two $x^2 - y^2$ type orbitals arranged in the same way as for Cu2 spins in volborthite shown in Fig. 1(d). In this case, the sign and magnitude of the interactions depend mostly on the bond angle. The Cu1-OH-Cu2 pathway must give an antiferromagnetic interaction with a magnitude similar to that of Cu2-OH-Cu2 because the angle is 101.85$^{\circ}$. On the other hand, the Cu1-O-Cu2 pathway must give a rather strong ferromagnetic interaction, because the Cu1-O-Cu2 angle is almost 90$^{\circ}$ (90.47$^{\circ}$).

To explain the small positive Weiss temperature of 13.4 K in KCu$_3$As$_2$O$_7$(OH)$_3$ within a nearest-neighbor coupling model on the kagome net, the Cu1-Cu2 magnetic interaction $J^{\prime}$ should be ferromagnetic; the Cu2-Cu2 interaction $J$ is undoubtedly antiferromagnetic. This means that the ferromagnetic Cu1-O-Cu2 superexchange pathway is dominant over the antiferromagnetic Cu1-OH-Cu2 pathway, resulting in the ferromagnetic $J^{\prime}$ in total. For example, if one assumes $J$ = 50 K, $J^{\prime}$ would be $-$45 K. Further neighbor interactions such as those via an As$^{5+}$ ion may exist, but they must be weakly antiferromagnetic.

Therefore, KCu$_3$As$_2$O$_7$(OH)$_3$ provides a unique distorted spin-1/2 kagome lattice with $J >$ 0 and $J^{\prime} <$ 0. Since $J$ between Cu2 spins is antiferromagnetic, magnetic couplings between Cu1 and Cu2 spins are always geometrically frustrated irrespective of the sign of $J^{\prime}$, unless $|J^{\prime}|$ is much larger than $J$. The peculiar magnetic order observed at $T_{\mathrm{N}}$ may be related to this frustration effect. Interesting frustration physics will be explored in the distorted kagome lattice of KCu$_3$As$_2$O$_7$(OH)$_3$ in future studies.

In summary, KCu$_3$As$_2$O$_7$(OH)$_3$ is found to be a spin-1/2 antiferromagnet with a highly distorted kagome lattice and to exhibit an unconventional antiferromagnetic order at $T_{\mathrm{N}}$ = 7.2 K. The distortion in magnetic interactions originates from a unique orbital arrangement comprising the 3$z^2 - r^2$ orbital in the Cu1 site and the $x^2 - y^2$ orbital in the Cu2 site. It is suggested that the magnetic interaction for Cu2-Cu2 is antiferromagnetic and that for Cu1-Cu2 is ferromagnetic. 


We thank S. Ishihara for helpful discussion. This work was partly supported by a Grant-in-Aid for Scientific Research on Priority Areas ``Novel States of Matter Induced by Frustration'' (No. 19052003) provided by the Ministry of Education, Culture, Sports, Science and Technology, Japan.


\end{document}